\documentstyle[12pt]{article}
\begin{document}
\setcounter{page}{1}
\pagestyle{plain}
\renewcommand{\theequation}{\arabic{section}.\arabic{equation}}
\setcounter{equation}{0}
%
%
%
\ \\[12mm]
\begin{center}
    {\bf Matrix Product Eigenstates for One-Dimensional Stochastic Models 
      \\[1mm] and Quantum Spin Chains}
             \\[20mm]
\end{center}
\begin{center}
\normalsize
	Klaus Krebs $^{\star}$ and
        Sven Sandow$^{\diamond}$\\[0.5cm]
		$^{\star}$ {\it Physikalisches Institut der Universit\"at Bonn\\
		 Nussallee 12, Bonn 53115, Germany}\\[4mm]
                $^{\diamond}$ {\it Department of Physics and Center for
                 Stochastic Processes in Science and Engineering,\\
                 Virginia Polytechnic Institute and State University\\
                 Blacksburg, VA 24061-0435, USA  }\\[1cm]

\end{center}
\vspace{2cm}
\begin{center}{\bf Abstract}\end{center}
We show that all zero energy eigenstates of an arbitrary $m$--state
quantum spin chain Hamiltonian with nearest neighbor interaction in the 
bulk and single site boundary terms, which can also describe the dynamics of
stochastic models, can be written as matrix product states. 
This means that the weights in these states can be expressed as expectation
values in a Fock representation of an algebra generated by $2m$ operators 
fulfilling $m^2$ quadratic relations which are defined by the Hamiltonian.
\\[20mm]
\rule{7cm}{0.2mm}
\begin{flushleft}
\parbox[t]{3.5cm}{\bf Key words:}
\parbox[t]{12.5cm}{Reaction-diffusion systems, quantum spin chains, matrix
product states
}
\\[2mm]
\parbox[t]{3.5cm}{\bf PACS numbers:} 02.50.Ey, 05.60.+w, 75.10.Jm, 82.20.Mj
\parbox[t]{12.5cm}{}
\end{flushleft}
\normalsize
\thispagestyle{empty}
\pagestyle{plain}

%
%
%
\newpage
\setcounter{page}{1}
\section{Introduction}
\setcounter{equation}{0}
\indent
A number of problems in many particle systems have been studied with the help
of so called matrix product states.
The idea of this technique is to express physical quantities such as ground
state wave functions or correlation functions as products of operators acting
on an auxiliary space and fulfilling algebraic relations defined by the
Hamiltonian of the system.
Introduced in the context of lattice animals \cite{hn} the technique has been
used to find ground states of various quantum spin chains 
\cite{brasil}-\cite{ksz}.

As it has been shown by Derrida, Evans, Hakim, and Pasquier \cite{dehp} the
steady state of the one species asymmetric exclusion process with open 
boundaries can be written as a matrix product state. 
Later works study this process in more detail \cite{sa}-\cite{der}.
Examples for asymmetric exclusion processes with two species were
investigated in \cite{djls,efgm} with the help of the same technique. 
The algebra used in these studies is generated by as many operators as states a
single site can take.
It has the important property that it leads to recurrence 
relations for the steady state of systems of different lattice lengths.
For other problems this algebra had to be generalized by 
enclosing additional operators.
This was first done in the study of the dynamics of the asymmetric exclusion
processe \cite{ss,gms}.
Another example is the algebra for the reaction--diffusion model studied in 
\cite{hsp} which is generated by twice as many operators as the one from
\cite{dehp}.
In this Fock--algebra the recursion property is lost. 
The same is true for the algebra used in \cite{haye}-\cite{hs} for stochastic
models with parallel updating.

One of the questions coming up naturally in this context is the following:
To which kind of problems can the matrix technique  be applied? 
In other words:
Does a Hamiltonian, which describes either a quantum spin system or a 
stochastic process, need to have any particular property in order to have
matrix product eigenstates?
In this paper we will prove the following proposition:
{\em Any zero energy eigenstate of a Hamiltonian with nearest neighbor
interaction in the bulk and single site boundary terms can be written as a 
matrix product state with respect to the Fock--algebra 
(\ref{bulk algebra})-(\ref{bc algebra}), which is that of Ref. \cite{hsp}.}
The technique of matrix product states was called matrix product ansatz in the
literature.
What we will see is that this technique, in the form of
(\ref{bulk algebra})-(\ref{bc algebra}), is not an ansatz but rather an 
identical reformulation of the eigenvector equation for the zero energy 
eigenstate.

The bulk of the paper is organized as follows: 
In Section 2 we define the class of Hamiltonians and in Section 3 we give the
proof of the proposition. 
We conclude with some remarks on matrix product techniques in Section 4.
%
%
\setcounter{equation}{0}
\section{Definition of a class of models }
\indent
The Hamiltonian we are going to consider in this paper is of the form
\begin{eqnarray}
\label{matrix ham}
H=h^{(L)}\;+\;
\sum_{j=1}^{N-1}  h_{j,j+1} \;+\;
 h^{(R)}
\end{eqnarray}
with
\begin{eqnarray}
\label{structure of h 1}
h^{(L)}&=&h^{(l)} \otimes I^{\otimes (N-1)}\;\;\;\;,\;\;\;\;
h^{(R)}= I^{\otimes (N-1)} \otimes h^{(r)}\\
\label{structure of h 2}
\mbox{and} \;\;\; h_{j,j+1}&=&
I^{\otimes (j-1)}\otimes h \otimes I^{\otimes (N-j-1)}
\end{eqnarray}
where $I$ is the $m \times m$-identity matrix, $h$ is an $m^2 \times m^2$-
matrix describing a two site interactions in the bulk,
and $h^{(l)}\;,\;h^{(r)}$ are $m\times m$-matrices defining single-site 
boundary terms.
This kind of Hamiltonian appears as well in the study of one--dimensional
stochastic systems as in the study of quantum spin chains.
We will now describe both applications in more detail.

{\em Stochastic model:} Let us consider a one-dimensional  lattice with $N$
sites, each of which can be in either of $m$ states. 
A configuration on the lattice is completely defined
by the set of occupation numbers $\{s_i\}=s_1,s_2,\ldots,s_N\;$
with $s_i=1,...,m\;\;\forall\;i=1,...,N\;$. The system evolves stochastically.
During an infinitesimal time step $dt$ its configuration $\{s_i\}$  can
change to a configuration $\{s'_i\}$ with a probability
$r(\{s_i\},\{s'_i\})dt$ where the $r(\{s_i\},\{s'_i\})$ are
referred to as rates. This process
can be described in terms of a rate equation which reads
\begin{equation}
\label{rate equation}
\partial_t P(\{s_i\},t) \;=\;
\sum_{ \{s'_j \ne s_i\}} \;\bigl[\;
r(\{s'_i\},\{s_i\})\;P(\{s'_i\},t)\;-\;
r(\{s_i\},\{s'_i\})\;P(\{s_i\},t)\;\bigr]
\end{equation}
where $P(\{s_i\},t)=P(s_1,s_2,\ldots,s_N,t)$ is the probability of finding the
configuration $s_1,s_2,\ldots,s_N$ at time $t\;$. Throughout this
paper we restrict ourselves  to dynamics where the configuration can change
only at two adjacent sites at a time and the rate for such a change depends
only on these two sites.
The rates are assumed to be independent of the position in the bulk of
system.
At the boundaries, i.e. at sites $1$ and $N$, we assume additional processes
to take place.

It is convenient to introduce a vector notation \cite{hsp} by writing
\begin{eqnarray}
\label{p as vector}
|P(t) \; ) \;=\;\sum_{ \{s_i\} }\;
P(s_1,s_2,\ldots,s_N,t) \;| s_1 \;) \otimes
| s_2 \;) \otimes...\otimes| s_N \;)
\end{eqnarray}
with
\begin{eqnarray}
\label{basis}
| k \;)\;=\;
\left( \hspace{-1.5mm} \begin{array}{c} 0 \\
.\\
.\\
0\\
1\\
0\\
.\\
.\\
0\end{array} \hspace{-1.5mm}  \right)
 \hspace{-1.5mm} \begin{array}{c} \leftarrow 1 \\
. \\
 .\\
. \\
\leftarrow k\\
 .\\
 .\\
 .\\
 \leftarrow m\end{array} \hspace{-1.5mm}
\;\;.\end{eqnarray}
In terms of these vectors the rate equation reads
\begin{eqnarray}
\label{vector rate equation}
\partial_t |P(t) \;)\;=\:-H |P(t) \;)
\;\;\end{eqnarray}
where $H$ is an $ m^N \times m^N$-matrix which is defined in
terms of the rates. We call $H$ the Hamiltonian here.
For the processes we are studying it has the structure of
(\ref{matrix ham})-(\ref{structure of h 2}).   
For stochastic models the matrices $H$ have a particular property:
They have to have vanishing column sums because the total probability has to
be conserved. 
This implies that there is a zero energy eigenstate for every lattice length $N$
which is just the steady state of the system.
It is denoted by $|P_{N} \;)$ and obeys the relation
\begin{eqnarray}
\label{eigenvector}
H \;|P_{N} \;)\;=0
\;\;.\end{eqnarray}
In Section 3 we will discuss a representation of these states.\\[0.2cm]

{\em Quantum spin chain:} Consider a chain of length $N$
with a spin $s$-particle sitting on each site. Suppose there is an interaction
between adjacent particles and some surface fields acting
on sites $1$ and $N\;$. Then the system dynamics is described by a
Schr\"odinger equation with a Hamiltonian of type
(\ref{matrix ham})-(\ref{structure of h 2}) if we chose a spin s-representation
and $m=2 s+1\;$.
The Hamiltonian generally does not have vanishing columns sums like
in the stochastic case,  though
it has to be hermitian. This condition does not effect the
construction of
matrix product states below.
Although the physical meaning of
the Schr\"odinger equation
is different from the meaning of the rate equation (\ref{vector rate equation})
it has the same  mathematical structure. And just as for the
stochastic model we are
often interested in the ground state of the  Hamiltonian $H$.
%
%
\section{Matrix product states}
\setcounter{equation}{0}
\indent
Let us now turn to the matrix product states \cite{dehp}-\cite{hsp}.
We introduce an auxiliary vector space $V_a$ and  we define
$2 m$ operators $D_s$ and $X_s$ with
$s=1,2,...,m$ acting on $V_a$ as well as
 two vectors
$|W >$ and $<V |$ in $V_a$. 
( Vectors in the auxiliary space
are denoted by $|...>$ in contrast to the vectors in the configuration space
which are denoted by $|...\;)\;$.) 
Next we define a vector
$|\tilde P_{N}\;)$ as
\begin{equation}
\label{eigenstate}
|\tilde P_{N}\;)\;=\; < W| \,
\left( \hspace{-1.5mm} \begin{array}{c} D_1 \\
 D_2\\
.\\
.\\
D_m\end{array} \hspace{-1.5mm}  \right) ^ {\otimes N}
|V >\,\,
\;\;.\end{equation}
where $\otimes$ stands for the direct product in the configuration space, 
so that the above equation reads in terms of its components:
$\;\tilde{P}_{N}(s_1,s_2,...,s_N)=
<W|D_{s_1}D_{s_2}...D_{s_{N-1}}D_{s_{N}}|V>\;$.
We call states of type $|\tilde P_{N}\;)$ matrix product states.

Let $H$ be a Hamiltonian of type (\ref{matrix ham})-(\ref{structure of h 2})
which has 
a zero energy eigenstate for all lattice lengths $N$ (a Hamiltonian describing
a stochastic model always has this property).
Then we can state the following proposition:\\[0.5cm]
{\bf (i)} {\em If $D_s\;,\;X_s$ and $|W >\;,\;<V |$ fulfill the following
relations}
\begin{equation}
\label{bulk algebra}
h\; \left[ \,
\left( \hspace{-1.5mm} \begin{array}{c} D_1  \\ D_2\\.\\.\\
 D_m \end{array} \hspace{-1.5mm} \right) \otimes
\left( \hspace{-1.5mm} \begin{array}{c} D_1  \\ D_2\\.\\.\\
D_m\end{array} \hspace{-1.5mm} \right)
\, \right] \;\;=\;\;
\left( \hspace{-1.5mm} \begin{array}{c} X_1\\ X_2\\.\\.\\
 X_m \end{array} \hspace{-1.5mm} \right) \otimes
\left( \hspace{-1.5mm} \begin{array}{c} D_1  \\ D_2\\.\\.\\
 D_m \end{array} \hspace{-1.5mm} \right) -
\left( \hspace{-1.5mm} \begin{array}{c} D_1  \\ D_2\\.\\.\\
 D_m \end{array} \hspace{-1.5mm} \right) \otimes
\left( \hspace{-1.5mm} \begin{array}{c} X_1\\ X_2\\.\\.\\
 X_m \end{array} \hspace{-1.5mm} \right)
\;\;,
\end{equation}
\begin{equation}
\label{bc algebra}
<W| \;h^{(l)}\; \left( \hspace{-1.5mm} \begin{array}{c}  D_1  \\ D_2\\.\\.\\
 D_m \end{array} \hspace{-1.5mm} \right) =
-<W| \, \left( \hspace{-1.5mm} \begin{array}{c} X_1\\ X_2\\.\\.\\
 X_m \end{array} \hspace{-1.5mm} \right)\,
\hspace{2mm}
\mbox{\em and}
\hspace{6mm}
h^{(r)}\; \left( \hspace{-1.5mm} \begin{array}{c}  D_1  \\ D_2\\.\\.\\
D_m \end{array} \hspace{-1.5mm} \right) \, |V> =
\left( \hspace{-1.5mm} \begin{array}{c} X_1\\ X_2\\.\\.\\
X_m \end{array} \hspace{-1.5mm} \right)\, |V>\;\;,
\end{equation}
{\em then  the vector $|\tilde P_{N}\;)$ defined by Eq. (\ref{eigenstate})
solves the equation for the zero energy eigenstate, i.e.,}
\begin{equation}
\label{eigenschlange}
 H\;|\tilde P_{N} \;)\;=0
\;\;.
\end{equation}
{\bf (ii)}
{\em For any vector
$|P_{N} \;)$ solving $H\;|P_{N} \;)=0$ 
one can find operators $D_s\;,\;X_s\;\; (s=1,2,...,m)$ and  vectors
$<W|\;,\;|V>$ in some space $V_a$, such that $|P_{N} \;)$
can be represented as a matrix product state $| \tilde P_{N} \;)$ defined by 
Eq. (\ref{eigenstate}) and relations (\ref{bulk algebra}) and 
(\ref{bc algebra}) are fulfilled in $V_a\;$.} \\[0.2cm]

Before we come to the proof of the statement we stress again that,
in contrast to the algebra defined in \cite{dehp}-\cite{er}, 
the relations (\ref{bulk algebra})-(\ref{bc algebra}) do not lead to recurrence
relations for expectation values of products of different length in the 
operators $D_s$ and $X_s$.   

The proof of (i)  is based on a
site by site cancellation of terms when $H$ is applied on 
$| \tilde P_{N} \;)\;$. 
The mechanism is basically the same as the one worked out in Ref. \cite{dehp}.
In Appendix A we explain it in detail.

In order to prove (ii) we construct a representation of operators and vectors 
fulfilling the Fock--algebra (\ref{bulk algebra})-(\ref{bc algebra}) using
the eigenvectors $|P_{N} \;)$ of systems with lengths $N=1,2,...\;$. 
Let us define for any number $M=1,2,...$ an $m^M$-dimensional space $V_M$
with a set of orthogonal  basis vectors $|s_1,s_2,...,s_M>\;(s_i=1,2,...,m)\;$
as well as  the one-dimensional space $V_0$ with the basis vector $|\;>\;$.
We define the space $V_a$ as the direct sum of all $V_M$ with
$M=0,1,2,...\;$.
The above basis vectors may be written as infinite column vectors with a $1$ 
at position  $1+s_1+m s_2+m^2 s_3+....+m^{M-1}s_M$
and a $0$ at all other positions, and their transposed $<s_1,s_2,...|$ can be 
written as the corresponding row vectors.
Next we define operators $D_s$ and $X_s$ as well as vectors $<W|$ and $|V>$ 
by means of their action on all the $|s_1,s_2,...>$ which provides a matrix 
representation if we write the $|s_1,s_2,...>$  as columns. 
The $D_s$ we define by
\begin{eqnarray}
D_{s_1}|s_2,s_3,...,s_{N}>
\label{matrix A}
\;\;\;\;\;\;=\;|s_1,s_2,s_3,..., s_{N}>\\
&&\;\;\;\mbox{for}\;\;N=1,2,...\;\;\mbox{and}\;\; s_i=1,2,...,m\nonumber
\;\;.\end{eqnarray}
The $X_s$ act on the basis vectors of $V_{N-1}\;$, i.e.\ on the 
$|s_1,s_2,...,s_{N-1}>\;$, as
\begin{eqnarray}
&&\left( \hspace{-1.5mm} \begin{array}{c} X_1\\ X_2\\.\\.\\
 X_m \end{array} \hspace{-1.5mm} \right) \otimes
\left( \hspace{-1.5mm} \begin{array}{c}
\overbrace{|1,1,...,1>}^{N-1 \;\;numbers}\\
|1,1,...,2>\\
.\\.\\.\\.\\.\\|m,m,...,m-1>\\|m,m,...,m>
\end{array} \hspace{-1.5mm} \right)\nonumber\\
&&\;\;\;\;\;\;=\;[h_{1,2}+h_{2,3}+...+h_{N-1,N}+h^{(R)}]\;
\label{matrix A bar}
\left( \hspace{-1.5mm} \begin{array}{c} D_1 \\D_2\\ .\\.\\
D_m\end{array} \hspace{-1.5mm}  \right) ^ {\otimes N}
\;|\; >\\
&&\;\;\;\mbox{for}\;\;N=1,2,...\nonumber
\;\;\end{eqnarray}
where $h_{i,i+1}$ and $h^{(R)}$
act on the N-fold tensor product in Eq. (\ref{matrix A bar}) according to
their definitions (\ref{structure of h 1}) and (\ref{structure of h 2}).
Note that each choice of $N$ gives a different
set of basis vectors in the column on the l.h.s.,
 and taking successively $N=1,2,...$  all basis
vectors will occur in this column exactly once. 
Hence, the above equation defines the action of each $X_s$ on
each basis vector in a consistent way.
The vectors $|V>$ and $<W|$ we define by
\begin{eqnarray}
\label{matrix V}
|V>&=&|\;>\\
\label{matrix W}
<W|s_1,s_2,...,s_N>&=&
P_{N}(s_1,s_2,....,s_N)\\
&&\mbox{for}\;\;N=1,2,...\;\;\mbox{and}\;\; s_i=1,2,...,m\nonumber
\end{eqnarray}
where $P_{N}(s_1,s_2,....,s_N)$ are the components of $|P_{N}\;)\;$. 
The above definitions fix the $D_s\;,\;X_s$ and the $<W|\;,\;|V>$ completely 
up to a constant $<W|\;>\;$. 
This constant can be fixed arbitrarily, since it does not enter into any result.

We have to prove now that these operators and
vectors  lead back to $|P_{N}\;)$ by means of Eq. (\ref{eigenstate}), i.e.,
that $|\tilde P_{N}\;)$  is equal to $|P_{N}\;)\;$,  and that they obey the 
Fock--algebra (\ref{bulk algebra})-(\ref{bc algebra}). 
Let us begin the first proof by rewriting (\ref{matrix A}) as
\begin{eqnarray}
\label{matrix A1}
D_{s_1}D_{s_{2}}...D_{s_N}|\;>=
|s_1,s_2,...,s_N>\;\\
\mbox{for}\;\;N=1,2,...\;\;\mbox{and}\;\; s_i=1,2,...,m\nonumber
\;\;.\end{eqnarray}
(The above relation gives a more intuitive meaning of the representation we have
chosen:
A basis vector $|s_1,s_2,...,s_M>$ is created by applying
the sequence  $D_{s_1}D_{s_2}...D_{s_M}$ on $|\;>\;$.)
Furthermore, using (\ref{matrix W}) we get:
\begin{eqnarray}
\label{W A1}
<W|D_{s_1}D_{s_{2}}...D_{s_N}|\;>=
P_{N}(s_1,s_2,....,s_N)\\
&&\mbox{for}\;\;N=1,2,...\;\;\mbox{and}\;\; s_i=1,2,...,m\nonumber
\;\;.\end{eqnarray}
Rewriting (\ref{W A1}) as a direct product of vectors in the configuration space
and using $|\;>=|V>$ yields
\begin{eqnarray}
\label{prove 1}
< W| \,\left( \hspace{-1.5mm} \begin{array}{c} D_1 \\D_2\\ .\\.\\
D_m\end{array} \hspace{-1.5mm}  \right) ^ {\otimes N} |V >\
\;=\;|P_{N} \;)\\
&&\mbox{for}\;\;N=1,2,...\nonumber
\;\;.\end{eqnarray}
which is what we were to prove.

In order to show that above representation fulfills the algebra
(\ref{bulk algebra})-(\ref{bc algebra}) we rewrite the
 definition (\ref{matrix A bar})
using Eq. (\ref{matrix A1}) and multiply the column vector
$\left( \hspace{-1.5mm} \begin{array}{c} {D_1}\\ {D_2}\\.\\.\\
{D_m} \end{array} \hspace{-1.5mm} \right) ^{\otimes n} $ from the left:
\begin{eqnarray}
&
\left( \hspace{-1.5mm} \begin{array}{c} {D_1}\\ {D_2}\\.\\.\\
{D_m} \end{array} \hspace{-1.5mm} \right) ^{\otimes n} \otimes
\left( \hspace{-1.5mm} \begin{array}{c} X_1\\ X_2\\.\\.\\
 X_m \end{array} \hspace{-1.5mm} \right) \otimes
\left( \hspace{-1.5mm} \begin{array}{c} {D_1}\\ {D_2}\\.\\.\\
{D_m} \end{array} \hspace{-1.5mm} \right) ^{\otimes (N-1-n)} |\;>\nonumber\\
&\;\;\;=\;[h_{n+1,n+2}+h_{n+2,n+3}+...+h_{N-1,N}+h^{(R)}]\;
\label{matrix A bar 1}
\left( \hspace{-1.5mm} \begin{array}{c} D_1 \\D_2\\ .\\.\\
D_m\end{array} \hspace{-1.5mm}  \right) ^ {\otimes N}
\;|\;>\\
&\;\;\;\mbox{for}\;\;n=0,1,...,N-1\;\;\mbox{and}\;\;N=1,2,...\nonumber
\;\;.\end{eqnarray}
The right equation of (\ref{bc algebra}) is nothing but the case $n=0\;,\;N=1$ 
of above relation. 
Consequently it holds. Next we use Eq. (\ref{matrix A bar 1}) to compute
\begin{eqnarray}
&&\left( \hspace{-1.5mm} \begin{array}{c} X_1\\ X_2\\.\\.\\
 X_m \end{array} \hspace{-1.5mm} \right) \otimes
\left( \hspace{-1.5mm} \begin{array}{c} {D_1}\\ {D_2}\\.\\.\\
{D_m} \end{array} \hspace{-1.5mm} \right) ^{\otimes (N-1)} |\;>
\;-\;
\left( \hspace{-1.5mm} \begin{array}{c} {D_1}\\ {D_2}\\.\\.\\
{D_m} \end{array} \hspace{-1.5mm} \right) \otimes
\left( \hspace{-1.5mm} \begin{array}{c} X_1\\ X_2\\.\\.\\
 X_m \end{array} \hspace{-1.5mm} \right) \otimes
\left( \hspace{-1.5mm} \begin{array}{c} {D_1}\\ {D_2}\\.\\.\\
{D_m} \end{array} \hspace{-1.5mm} \right) ^{\otimes (N-2)} |\;> \nonumber\\
\label{prove 1a}
&&\;\;\;\;=\;h_{1,2} \;
\left( \hspace{-1.5mm} \begin{array}{c} {D_1}\\ {D_2}\\.\\.\\
{D_m} \end{array} \hspace{-1.5mm} \right) ^{\otimes N} |\;>
\;\;,\end{eqnarray}
i.e.,
\begin{eqnarray}
&&\left \{\;
\left( \hspace{-1.5mm} \begin{array}{c} X_1\\ X_2\\.\\.\\
 X_m \end{array} \hspace{-1.5mm} \right) \otimes
\left( \hspace{-1.5mm} \begin{array}{c} {D_1}\\ {D_2}\\.\\.\\
{D_m} \end{array} \hspace{-1.5mm} \right) \;-\;
\left( \hspace{-1.5mm} \begin{array}{c} {D_1}\\ {D_2}\\.\\.\\
{D_m} \end{array} \hspace{-1.5mm} \right) \otimes
\left( \hspace{-1.5mm} \begin{array}{c} X_1\\ X_2\\.\\.\\
 X_m \end{array} \hspace{-1.5mm} \right)\;-\;
h\;
\left[ \,
\left( \hspace{-1.5mm} \begin{array}{c} D_1  \\ D_2\\.\\.\\
 D_m \end{array} \hspace{-1.5mm} \right) \otimes
\left( \hspace{-1.5mm} \begin{array}{c} D_1  \\ D_2\\.\\.\\
D_m\end{array} \hspace{-1.5mm} \right)
\, \right] \;
\right \}\;\otimes\nonumber\\
\label{prove 2}
&&\left( \hspace{-1.5mm} \begin{array}{c} {D_1}\\ {D_2}\\.\\.\\
{D_m} \end{array} \hspace{-1.5mm} \right) ^{\otimes (N-2)} |\;>\;=\;0\\
&&\;\;\;\mbox{for}\;\;N=2,3,...\nonumber
\;\;.\end{eqnarray}
Since the products  of the $D_s$ span the whole space $V_a\;$,
Eq. (\ref{bulk algebra}) must hold in  $V_a\;$. What remains to be
proven  is the left  equation of (\ref{bc algebra}). We multiply
Eq. (\ref{matrix A bar 1}) for $n=0$ by $<W|$ from the left
 and use Eq. (\ref{prove 1}):
\begin{eqnarray}
&&<W|
\left( \hspace{-1.5mm} \begin{array}{c} X_1\\ X_2\\.\\.\\
 X_m \end{array} \hspace{-1.5mm} \right) \otimes
\left( \hspace{-1.5mm} \begin{array}{c} {D_1}\\ {D_2}\\.\\.\\
{D_m} \end{array} \hspace{-1.5mm} \right) ^{\otimes (N-1)} |\;> \nonumber\\
\label{prove 3}
&&\;\;\;\;\;=[h_{1,2}+h_{2,3}+...+h_{N-1,N}+h^{(R)} ]\;|P_{N}\;)
\;\;.\end{eqnarray}
Writing now the Hamiltonian of an $N$-site system as
$H=h^{(L)}+h_{1,2}+h_{2,3}+...+h_{N-1,N}+h^{(R)}$ and using
$H|P_{N}\;)=0\;$, the above relation can be written as:
\begin{eqnarray}
\label{prove 4}
&&<W|
\left( \hspace{-1.5mm} \begin{array}{c} X_1\\ X_2\\.\\.\\
 X_m \end{array} \hspace{-1.5mm} \right) \otimes
\left( \hspace{-1.5mm} \begin{array}{c} {D_1}\\ {D_2}\\.\\.\\
{D_m} \end{array} \hspace{-1.5mm} \right) ^{\otimes (N-1)} |\;>
\;=\;-h^{(L)}\;|P_{N}\;)
\;\;.\end{eqnarray}
Using again Eq. (\ref{prove 1}) yields
\begin{eqnarray}
\label{prove 5}
<W|\;\left \{\;
h^{(l)} \left( \hspace{-1.5mm} \begin{array}{c} {D_1}\\ {D_2}\\.\\.\\
{D_m} \end{array} \hspace{-1.5mm} \right) \;+\;
\left( \hspace{-1.5mm} \begin{array}{c} X_1\\ X_2\\.\\.\\
 X_m \end{array} \hspace{-1.5mm} \right)\;
\right \}\;\otimes
\left( \hspace{-1.5mm} \begin{array}{c} {D_1}\\ {D_2}\\.\\.\\
{D_m} \end{array} \hspace{-1.5mm} \right) ^{\otimes (N-1)} |\;>\;=\;0\\
\;\;\;\mbox{for}\;\;N=1,2,...\nonumber
\;\;.\end{eqnarray}
Consequently the left equation of  (\ref{bc algebra}) holds in $V_a\;$. 
This completes the proof that the  matrices defined by Eq. 
(\ref{matrix A})-(\ref{matrix W}) represent the algebra
(\ref{bulk algebra})-(\ref{bc algebra}) and hence the proof of
proposition (ii). \\[0.3cm]

Let us add some remarks on the proof. 
We proved that the Fock algebra (\ref{bulk algebra})-(\ref{bc algebra}) has a
nontrivial representation if an eigenstate with zero energy
exists for all lattice lengths.
The construction of the matrices obeying (\ref{bulk algebra}) can be done 
without any demand for special properties of $H$.
(An associative algebra generated by $2m$ generators with $m^2$ quadratic 
relations is always a well defined mathematical object.)
The special property of $H$, i.e.\ the existence of zero energy eigenstates 
$|P_{N}(s_1,s_2,...,s_N))$, was only needed for the construction of a non 
trivial scalar product involving the vectors $<W|$ and $|V>$.
More precisely, the vectors $|P_{N}(s_1,s_2,...,s_N))$ only enter the 
definition of the vector $<W|$ and the equation which determines 
$<W|$ is just the equation for the zero energy eigenstate.

Due to the lack of recurrence relations the scalar product in 
(\ref{bulk algebra})-(\ref{bc algebra}) is not completely fixed and can be 
chosen independently in each vector space $V_M$. 
For that reason it would be sufficient to require that at least for one lattice 
length $N$ the Hamiltomian has a zero energy eigenstate.
In that case $<W|$ is nontrivial only on the subspace $V_N$.

One may define a finite dimensional $V_a$ as the direct sum over all $V_M$
with $M \le N_{max}\;$ for some number $N_{max}\;$. A matrix representation of 
the algebra (\ref{bulk algebra})-(\ref{bc algebra}) can then be found
be assuming Eq. (\ref{matrix A})-(\ref{matrix W}) for $N=1,2,...,N_{max}$ as 
well as $D_s|s_1,s_2,...,s_{N_{max}}>=0$ and 
$X_s|s_1,s_2,...,s_{N_{max}}>=0\;$.
Then Eq. (\ref{eigenstate}) provides a matrix product representation for all 
eigenvectors $|P_N\;)$ with $N\le N_{max}\;$.

It is of course also possible  to express eigenstate of non zero eigenvalues 
as matrix product state by adding an appropriate shift to the Hamiltonian.
This is done by replacing $h$, $h^{(l)}$ and $h^{(r)}$ in Eqs. 
(\ref{bulk algebra})-(\ref{bc algebra}) by $h-\epsilon$, 
$h^{(l)}-\epsilon^{(l)}$ and $h^{(r)}-\epsilon^{(r)}$, respectively.
In this case the Hamiltonian $H$ in Eq. (\ref{eigenschlange}) has to be 
replaced by $H-E(N)$ with $E(N)=\epsilon^{(l)}+\epsilon^{(r)}+(N-1)\epsilon$ if 
$N$ is the lattice length.
The vector $|P_N)$ is then an eigenstate with energy $E(N)$.
However, if $\epsilon^{(l)},\epsilon^{(r)}$ and $\epsilon$ are independent 
of $N$, Eq. (\ref{eigenschlange}) will have a non trivial solution only for
a finite number of lattice lengths $N$ and $<W|$ will therefore be
non trivial only on the corresponding subspaces $V_N$ of $V_a$.
%
%
%
\section{Concluding Remarks}
\indent
We proved that any zero energy eigenstate of a Hamiltonian of the form 
(\ref{matrix ham}) can be written as a matrix product state with respect to 
the Fock representation of the algebra (\ref{bulk algebra})-(\ref{bc algebra}).
The proof has been done by showing that the operators as well as the scalar 
product can be defined in a non trivial way.
The equations defining the scalar product are exactly the equations for the 
zero energy eigenstate.
This shows that by application of the algebra on an abstract level, 
i.e.\ by using only the relations (\ref{bulk algebra})-(\ref{bc algebra}), 
one can not gain any insight into the form of the eigenstates.
One ends up with nothing but a reformulation of the equation for the 
zero energy eigenstate.
Therefore the so called {\em matrix product ansatz}, i.e.\ the application of 
the Fock--algebra (\ref{bulk algebra})-(\ref{bc algebra}) to an eigenvector 
problem, is not really an ansatz, it is an identity.
The eigenvector problem is not transformed into another problem,
it remains unchanged.
However, the situation becomes different if one considers a more special form
of the algebra as done in \cite{dehp}-\cite{efgm} where the operators $X_s$ are
replaced by numbers.

We checked that our proposition is also true for the Fock--algebra describing
stochastic models with parallel updating \cite{haye}.

%
%
%
%
\noindent
{\bf Acknowledgments}\\[1mm]
We thank H. Simon,
M. R. Evans, T. Heinzel, H. Hinrichsen, A. Honecker, and especially 
V. Rittenberg for valuable discussions. S.S. gratefully acknowledges financial 
support by the Deutsche Forschungsgemeinschaft.
%
%
%
%
\renewcommand{\thesection}{\Alph{section}}
\setcounter{section}{0}
\section{Appendix: Proof of eigenvalue equation for matrix product
state }
\renewcommand{\theequation}{\Alph{section}.\arabic{equation}}
\setcounter{equation}{0}
\indent
Following the line of Ref. \cite{dehp}
we prove proposition (i), i.e., we
show that the states (\ref{eigenstate}) solve the eigenvalue equation
(\ref{eigenschlange})
if (\ref{bulk algebra})-(\ref{bc algebra}) are fulfilled. 
Using the definition (\ref{matrix ham})  we write
\begin{eqnarray}
H&=&h^{(l)}\; \otimes I^{\otimes (N-1)}\;+\;
\sum_{j=1}^{(N-1)} I^{\otimes (j-1)} \otimes h\otimes I^{\otimes (N-j-1)}
\nonumber\\
\label{H-E}&&\;\;\;\;
+\;I^{\otimes (N-1)} \otimes  h^{(r)}
\end{eqnarray}
Next we define a state $|\overline P_{N}(k) \;)$ as
\begin{equation}
\label{pbar}
|\overline P_{N}(k)\;) \;=\; < W| \,
\left( \hspace{-1.5mm} \begin{array}{c} D_1 \\
 D_2\\
.\\
.\\
D_m\end{array} \hspace{-1.5mm}  \right) ^ {\otimes (k-1)} \otimes
\left( \hspace{-1.5mm} \begin{array}{c} X_1 \\
 X_2\\
.\\
.\\
X_m\end{array} \hspace{-1.5mm}  \right) \otimes
 \left( \hspace{-1.5mm} \begin{array}{c} D_1 \\
 D_2\\
.\\
.\\
D_m\end{array} \hspace{-1.5mm}  \right) ^ {\otimes (N-k)}
|V >\,\,.
\end{equation}
Applying now the terms occurring in Eq. (\ref{H-E}) on the state
\begin{equation}
\label{eigenstate1}
|\tilde P_{N}\;)\;=\; < W| \,
\left( \hspace{-1.5mm} \begin{array}{c} D_1 \\
 D_2\\
.\\
.\\
D_m\end{array} \hspace{-1.5mm}  \right) ^ {\otimes N}
|V >\,\, \nonumber
\end{equation}
yields
\begin{eqnarray}
\label{hl on p}
h^{(l)}\; \otimes I^{\otimes (N-1)}\; |\tilde P_{N}\;)\;&=&\;
-|\overline P_{N}(1) \;)\\
\label{h on p}
I^{\otimes (j-1)} \otimes h\otimes I^{\otimes (N-j-1)}\; 
|\tilde P_{N}\;)\;&=&\;
|\overline P_{N}(j) \;)\;-\;|\overline P_{N}(j+1) \;)\\
\label{hr on p}
I^{\otimes (N-1)} \otimes h^{(r)}\; |\tilde P_{N}\;)\;&=&\;
|\overline P_{N}(N) \;)
\;\;\end{eqnarray}
where the algebra (\ref{bulk algebra})-(\ref{bc algebra}) was used. 
Applying now the full operator $H$ (see Eq. (\ref{H-E})) on the state 
$|\tilde P_{N}\;) $ we get
\begin{eqnarray}
H\;|\tilde P_{N}\;)\;&=&\;-|\overline P_{N}(1) \;)\;+\;
\sum_{j=1}^{N}\;(\;
|\overline P_{N}(j) \;)\;-\;|\overline P_{N}(j+1) \;)\;)
\;+\;|\overline P_{N}(N) \;) \nonumber\\
\label{H-E on p}
&=&\;0
\end{eqnarray}
which is what we were to show.
%
%

\end{document}